\documentclass[]{aa}

\def\eps@scaling{.95}
\def\epsscale#1{\gdef\eps@scaling{#1}}

\def\plotone#1{\centering \leavevmode
    \epsfxsize=\eps@scaling\columnwidth \epsfbox{#1}}

\def\plotfiddle#1#2#3#4#5#6#7{\centering \leavevmode
    \vbox to#2{\rule{0pt}{#2}}
    \includegraphics{#1}}

\usepackage{graphicx}
\usepackage{subfigure}
\begin{document}
\thesaurus{09.04.1; 11.09.1; 11.19.2; 11.19.5; 13.09.1; 13.20.1}
\title{Modelling the spectral energy distribution of galaxies.
I. Radiation fields and grain heating in the edge-on spiral NGC891
}
\author{Cristina C. Popescu \inst{1,}\thanks{Research Associate at 
The Astronomical Institute of the Romanian Academy, Str. Cu\c titul de Argint 
5, 75212, Bucharest, Romania}
\and  Angelos Misiriotis  \inst{2,3}
\and Nikolaos D. Kylafis \inst{2,4}
\and Richard J. Tuffs \inst {1}
\and J\"org Fischera \inst {1}
}

\offprints{Cristina C. Popescu (email address: Cristina.Popescu@mpi-hd.mpg.de)}

\institute{Max Planck Institut f\"ur Kernphysik, Saupfercheckweg 1, 
           D--69117 Heidelberg, Germany
\and University of Crete, Physics Department, P.O. Box 2208, 710 03
   Heraklion, Crete, Greece
\and Observatoire de Marseille , 2 place Le Verrier 
13248 Marseille Cedex 4, France
\and Foundation for Research and Technology-Hellas, P.O. Box 1527, 711 10 
   Heraklion, Crete, Greece
}

\date{Received; accepted}
\maketitle
\markboth{C.C.Popescu et al.}{Radiation field and grain heating in NGC891}
\begin{abstract}
We describe a new tool for the analysis of the UV to the sub-millimeter 
(sub-mm) spectral energy distribution (SED) of spiral galaxies. We use a 
consistent treatment of grain heating and emission, solve 
the radiation transfer problem for a finite disk and bulge, and 
self-consistently calculate the stochastic heating of grains placed in the
resulting radiation field.

We use this tool to analyse the well-studied nearby edge-on spiral galaxy 
NGC891.  First we investigate whether the old stellar population in NGC891,
along with a reasonable assumption about the young stellar population, can
account for the heating of the dust and the observed far-infrared and
sub-mm emission.  The dust distribution is taken from the model of
Xilouris et al. (1999), who used only optical and near-infrared observations
to determine it.  We have found that such a simple model cannot reproduce the
SED of NGC891, especially in the sub-mm range. It underestimates by a factor 
of 2-4 the observed sub-mm flux.

A number of possible explanations exist for the missing sub-mm flux.  We
investigate a few of them and demonstrate that one can reproduce the observed
SED in the far-infrared and the sub-mm quite well, as well as the observed
radial profile at $850\,\mu$m.

For the models calculated we give the relative proportion
of the dust radiation powered by the old and young stellar populations as
a function of FIR/sub-mm wavelength. In all models we find that the dust
is predominantly heated by the young stellar population.

\end{abstract}

\keywords{dust, extinction - galaxies: individual: NGC891 - galaxies: spiral - galaxies: stellar 
content - infrared: galaxies - submillimeter }

\section{Introduction} 

Dust grains can be considered as test particles for the intrinsic radiation 
field in galaxies. Observations of their emission in the infrared (IR), 
combined with optical and ultraviolet
(UV) observations of the light from stars, attenuated
by the grains, should, in principle, strongly constrain the intrinsic
distribution of stellar luminosity and dust in galaxies.  This would address 
the fundamental question of optical thickness in galactic disks and allow
evaluation of optical and IR observational data in terms of physical 
(e.g., star formation history) rather
than empirical (e.g., colours) parameters. Ultimately, such analysis techniques
will be necessary to interpret the evolution of the 
UV - far-infrared (FIR)/sub-millimeter (sub-mm) spectral energy distribution 
(SED) of galaxies and the IR background radiation over cosmological
timescales. 

However, even for galaxies in the local universe, technical difficulties are 
considerable from both theoretical and
observational stand points. From the theoretical point of view, it is necessary
to solve the radiation transfer problem for a finite disk, and self
consistently calculate the stochastic heating of grains placed in the
resulting radiation field using realistic grain models. To that must be added
the problem that disks are fundamentally inhomogeneous on different
scales. From the observational point of view, it is only now becoming possible
to obtain resolved images covering the mid-IR to sub-mm spectral range for
nearby galaxies.

Because of these difficulties, the origin of the FIR emission 
in local universe spiral galaxies is still a subject of 
debate. Based on 
models of dust heating in various environments (e.g., Mezger et al. 1982; 
Cox et al. 1986), and from comparison of IRAS data 
with optical properties of galaxies, several studies concluded that a 
substantial fraction of the FIR emission from galaxies could be due to dust 
heated by the diffuse interstellar radiation field (ISRF), the remaining part 
being due to localised sources around OB stars in star-forming regions 
(e.g., de Jong et al. 1984; Walterbos \& Schwering 1987). Some later papers, 
however, suggested that the correlation
between H${\alpha}$ and FIR emission implies that OB stars by themselves can
account for all the dust heating, so that  there is no need for an additional
contribution to heating by the ISRF (e.g., Devereux \& Young 1990). By
contrast, there have been a series of papers (e.g., Xu et al. 1994) which 
concluded that the non-ionising UV accounts for most of the
energy absorbed by dust. They showed that the FIR/radio correlation can be 
separated into a warm-FIR/thermal-radio correlation which is due to massive 
ionising stars ($>20\,{\rm M}_{\odot}$), and a cool-FIR/nonthermal-radio 
correlation, principally due to intermediate mass stars 
($5-20\,{\rm M}_{\odot}$) - the supernova 
progenitors. A heating component from the old stellar population was also 
needed to account for the non-linearity of the correlation, but this 
accounts only for about a third of the heating.

In this paper we describe a new method for the analysis of the radiation energy
budget in edge-on spiral galaxies, as part of an effort to develop a 
theoretical tool for the interpretation of the SED in spiral galaxies. The
method presented here is based on some knowledge of the primary 
radiation field (i.e., from stars) in spiral galaxies and the dust 
distribution in them.  For edge-on spirals, it has been demonstrated 
(Xilouris et al. 1997; 1998; 1999) that it is possible to determine the 
distribution of older stars and associated dust from analysis of optical and
near-infrared (NIR) images. 
Thus, the diffuse radiation field in the galaxy from 
these optical bands can be computed quite accurately. The diffuse radiation 
field in the UV is generally not so well constrained by the UV  data alone, 
due to the larger optical depth of disks in the UV. This younger stellar 
component is 
parameterised in terms of a recent star-formation rate (SFR) and it can be 
observationally constrained in the case that a direct, 
optically thin indicator, such as radio free-free emission, is available. 
Solving the radiation transfer problem within the disk and bulge, and
calculating the heating of grains we can determine the diffuse FIR/sub-mm
emission. Potentially, further contributions arising from emission from 
localised sources need to be added to the diffuse component.
Then, because the dust is optically thin in the FIR, the predicted 
emission can be compared with observations. 

We apply the above method to the well-known edge-on spiral galaxy NGC891, for 
which we assume a distance of 9.5 Mpc (van der Kruit \& Searle 1981). This 
is one of the most extensively observed edge-on galaxies in the nearby 
universe,
which makes it ideal for a verification of our modelling technique. 
The analysis of a sample of 6 more edge-on galaxies is in progress 
(Misiriotis et al. 2000a).
Modelling edge-on spiral
galaxies has several advantages, mainly when investigating 
them in the optical band. One advantage is that, in this view of a galaxy, 
one can easily
separate the three main components of the galaxy (i.e., the stellar disk, the
dust and the bulge). Another advantage is that the dust is very prominently 
seen in the dust lane, and thus its scalelength and scaleheight can be better 
constrained. A third advantage is that many details of a galaxy that are 
evident when the galaxy is seen face-on (e.g., spiral arms), are smeared out 
to a large degree when the galaxy is seen edge-on (Misiriotis et al. 2000b). 
Thus, a simple model with relatively
few parameters can be used for the distribution of stars and dust in the
galaxy. However, the third advantage comes with disadvantages, especially when
trying to model the galaxy in all wavelength ranges, including UV and
FIR/sub-mm. Thus, in edge-on galaxies it is very difficult to see
localised sources (i.e., HII regions), in which the radiation can be locally
absorbed and thus not contribute to the diffuse radiation field. Also, if the
galaxy has a thin (young) stellar/dust disk, highly obscured by the dust lane 
in the plane of the galaxy, then this disk cannot be inferred from 
observations in the optical/NIR spectral range. In passing we mention that
throughout this paper we will use the terms ``thin/thick disks'' to describe
their scaleheights, and not their optical thickness. 

The goal of this paper is to find the simplest phenomenological description of
NGC891 that can adequately account for the observed optical to sub-mm SED. In
particular we investigate what fraction of the FIR is accounted for
by the Xilouris
et al. (1999) model, namely by a disk of dust, heated by a diffuse, smooth,
optically emitting, old stellar disk and bulge, 
supplemented by a diffuse, smooth,
UV emitting disk of newly-formed stars. We refer to this model as the 
\lq\lq standard model\rq\rq.  We will show that this model fails
to reproduce the observed SED of the galaxy at FIR/sub-mm wavelengths. 
We investigate then different possibilities to explain the origin of the 
missing FIR/sub-mm flux.

Recently there have been several works
modelling the SED of galaxies from the UV to the 
sub-mm (Silva et al. 1998, Devriendt et al. 1999). Their method consists of
using models for photometric and/or spectrophotometric and chemical 
evolution of galaxies in order to fit the observed SED. For example Silva et
al. (1998) used a chemical evolution code to follow the SFR, the gas 
fraction, and the metallicity of the galaxies. In their approach the
parameters describing the star formation history as well as the geometrical
parameters of the intrinsic distribution of stars and dust are left as 
fitting parameters. While this approach was very successful in fitting 
the SED of different types of galaxies, it implies the use
of many free parameters. Furthermore, a more detailed description of the 
geometry of the galaxy is indispensable for an accurate determination of 
the dust content in individual cases. 
   
This paper is organised as follows:
In Sect. 2 we describe our ``standard model'' for the diffuse dust and 
radiation field,
including the optical properties of the dust grains, and the derivation of the
diffuse radiation field powered by the old and the young stellar populations.
In Sect. 3 we give the results of our calculations, in
terms of integrated SED for the \lq\lq standard model\rq\rq\, and in
Sect. 4 we supplement our model with the contribution of localised sources.
 Since the \lq\lq standard model\rq\rq\, fails to reproduce  the observed 
SED, we discuss in Sect. 5 the origin of the FIR emission, considering four
possibilities to produce the missing IR component.  From these, the two dust 
disk model, together with the clumpy scenario constitute the simplest 
solutions we have been able to identify to describe the origin of the missing 
infrared component. In Sect. 6 we give our
results in terms of radial profiles and in Sect. 7 we discuss the contribution
of radiation at different wavelengths in heating the dust. In Sect. 8 we
summarise our results and give our conclusions.

\section{The model}

\subsection{The properties of the dust grains}
We consider only the spectral range $\lambda \ge 40\,\mu$m as our dust model
does not include 
Polycyclic Aromatic Hydrocarbon (PAH) molecules, which are important for 
$\lambda \le 15\,\mu$m and may even contribute at $\lambda = 25\,\mu$m, a point
that is still very uncertain. For 
example, D\'esert et al. (1990) proposed a dust model in which 48\% of the 
IR emission at 25\,${\mu}$m (for dust in the solar neighbourhood) is emitted 
by the PAH molecules as continuum emission. 
Draine and Anderson (1985) proposed that 
the IR emission at 25\,${\mu}$m and shorter wavelengths
can be explained by
enhancing the number of small grains. Our restriction to the spectral range 
$\lambda \ge 60\,\mu$m allows us to avoid the uncertainties introduced by a
different dust component (PAHs or modified grain size distribution).

In characterising the dust properties we thus considered spherical \lq\lq 
astronomical grains\rq\rq\, for two component materials - graphite and 
silicate - of the Mathis et al. (1977; hereafter MRN) 
interstellar grain model. MRN have proposed that the observed interstellar
extinction of star-light may be produced by a mixture of graphite and silicate
particles with a simple power-law distribution of sizes

\begin{eqnarray}
dN_i = N_i(a) \, da = N_i\,a^{-k}\,da
\end{eqnarray}\\
where $dN_i$ is the number of grains of type $i$ with radii in the
interval $[a,a+da]$, and $k=3.5$. We consider the 
lower and the upper cutoffs to the size distribution to be respectively
$a_{\rm min} = 10\,\rm \AA$ and $a_{\rm max} = 0.25\,{\mu}$m.  The 
graphite and silicate abundances were taken from Draine \& Lee (1984, 
hereafter DL), and are only slightly different from those found by MRN; DL 
proposed a mixture of 53\% silicates ($N_{\rm Si}$) and 47\% graphites 
($N_{\rm Graphite}$), which were chosen to fit the extinction curve in our
Galaxy, and which we also adopted for NGC891. 
  
The absorption efficiencies $Q_{\rm abs}({\lambda},a)$, the scattering 
efficiencies $Q_{\rm scat}({\lambda},a)$ and the scattering phase function
$g({\lambda},a)$  were taken from Laor \& Draine (1993). The absorption 
coefficient ${\kappa}_{\rm abs}$, the scattering coefficient
${\kappa}_{\rm scat}$ as well as the averaged anisotropy parameter of the 
Henyey-Greenstein scattering phase function $g$ can then be
derived by integrating $Q_{\rm abs}({\lambda},a)$, 
$Q_{\rm scat}({\lambda},a)$, and $g({\lambda},a)$, respectively, over the size
distribution
\begin{eqnarray}
{\kappa}_{\rm abs}({\lambda}) & \sim & \int_{a_{\rm min}}^{a_{\rm max}}\sum_i
N_{\rm i}\,a^{\rm -k}da\,{\pi}\,a^2\,Q_{\rm abs}({\lambda},a)\\
{\kappa}_{\rm scat}({\lambda})& \sim & \int_{a_{\rm min}}^{a_{\rm max}}\sum_i
N_{\rm i}\,a^{\rm -k}da\,{\pi}\,a^2\,Q_{\rm scat}({\lambda},a)\\
g(\lambda) & = & \frac{\displaystyle \int_{a_{\rm min}}^{a_{\rm max}}
\sum_i N_{\rm i}\,a^{\rm -k}da\,{\pi}\,a^2\,g(\lambda, a) Q_{\rm scat}
({\lambda},a)}{\displaystyle \int_{a_{\rm min}}^{a_{\rm max}}\sum_i N_i
\,a^{-k}da\,{\pi}\,a^2\,Q_{\rm scat}({\lambda},a)}
\end{eqnarray}\\
with the extinction coefficient ${\kappa}_{\rm ext} =
{\kappa}_{\rm abs}+{\kappa}_{\rm scat}$.  

As the $60-100\,\mu$m data are expected to contain a significant amount of
emission from small grains not in equilibrium with the radiation field, it is
necessary to model the stochastic emission from the dust. For this calculation
we adopted the heat capacities of silicate grains from Guhathakurta \& Draine 
(1989), which were derived as a fit to experimental results for 
SiO$_2$ and obsidian at temperatures $10<T < 300$\,K (Leger et al. 1985), 
with a simple extrapolation for $T>300$\,K.  For 
graphite grains the heat capacities were taken from Dwek (1986).

\subsection{Derivation of the diffuse optical and UV radiation field.}
Our first
step is to estimate the diffuse radiation field at any point in the galaxy. 
The most straightforward way to do this is to solve the radiative transfer 
equation (RTE) for a given distribution of emitters and absorbers of 
radiation. In this section we will consider the distribution of emitters 
and absorbers as derived from optical wavelengths by Xilouris et
al. (1998,1999), while at the UV wavelengths we will fix the parameters
describing the geometry of the emitters and use population synthesis 
models to parameterise the amplitude of the UV radiation. 

\subsubsection{Radiation transfer} 
In this subsection we will describe our solution of the RTE. Our method
is based on that of Kylafis \& Bahcall (1987). 
This method will be briefly described here
for the sake of completeness. Let us introduce the RTE (Mihalas 1978) for
a time independent radiation field along a line of sight, at a fixed
wavelength 
\begin {equation}
\frac{dI(s,\hat{n})}{ds}=-\kappa_{ext}\,I(s,\hat{n}) + 
\eta_0(s,\hat{n}) + \eta_s(s,\hat{n})
\end {equation}
where $I(s,\hat{n})$ is the specific intensity of the radiation field 
at a point $s$
on the line of sight (defined by $\hat{n}$), $\kappa_{ext}$ is the 
extinction coefficient for the wavelength considered,
$ds$ is an infinitesimal length element along the line of sight,
$\eta_0$ is the emissivity
along the line of sight due to primary sources (stars) 
and $\eta_s$ is the 
emissivity due to scattering into the line of sight. 
For coherent scattering, $\eta_s$ is given by
\begin {equation}
\eta_s(s,\hat{n})=
\omega\,\kappa_{ext}
\int{I(s,\hat{n}')\,p(\hat{n},\hat{n}')\,
\frac{d{\Omega}'}{4\pi}}
\end {equation}
where $p(\hat{n},\hat{n}')$ is the Henyey-Greenstein phase function 
(Henyey \& Greenstein 1941), parameterised by the anisotropy parameter $g$,
and $\omega$ is the scattering albedo for the wavelength considered, 
defined as $\omega \equiv \kappa_{scat}/\kappa_{ext}$.
Let $I=I_0+I_1+...+I_n +...$, where $I_0$ is the intensity of photons that come 
directly from the primary sources, $I_1$ the intensity of photons 
that have been 
scattered once after leaving the primary sources and $I_n$ the intensity 
of photons that have suffered $n$ scatterings. The intensities
$I_n$ satisfy the following equations (Henyey 1937)
\begin{equation}
\frac{dI_n(s,\hat{n})}{ds}= -\kappa_{ext}\,I_n(s,\hat{n})+\eta_n(s,\hat{n})
\end{equation}
where $\eta_n(s,\hat{n})$ is the emissivity due to photons scattered $n$ 
times, and is given by
\begin{equation}
\eta_n(s,\hat{n})
=\omega\kappa_{ext}
\int {I_{n-1}(s,\hat{n}')\,p(\hat{n},\hat{n}')}\,\frac{d{\Omega}'}{4\pi}
\end{equation}
Eq. (7) can be solved for as high order of $n$ as allowed by the given
hardware and/or software. In our method, we calculate $I_0$ and $I_1$
and approximate the higher orders according to Kylafis \& Bahcall
(1987). Therefore, the solutions are given by
\begin{equation}
I_{0,1}(s,\hat{n})
=\int_0^s {ds'\,\eta_{0,1}(s',\hat{n})\,{\rm exp}[-\tau(s,s')]}
\end{equation}
while for $n>1$,
\begin {equation}
I_{n+1}=I_n \frac{I_1}{I_0}
\end {equation}
where $\tau(s,s')$ is the optical depth between $s$ and $s'$.
Given the intensity of the radiation field at a point, the energy density
can be calculated by (Mihalas 1978)
\begin{equation}
u(x,y,z)=c^{-1}\int{I(x,y,z,\hat{n})\,d\Omega}
\end{equation}

\subsubsection{Optical radiation field and dust distribution} 
The determination of the intrinsic stellar content of a galaxy is often 
difficult because of the uncertainty introduced by dust obscuration.
However, in the case of edge-on galaxies, dust obscuration can be used
to our benefit to determine the spatial distribution of dust. In an
edge-on galaxy, dust becomes visible as a very prominent dust lane
along the major axis of the galaxy. The effects of the dust are 
strong enough to allow the existence of a fairly constrained model 
for the distribution of old stars and associated dust 
(Xilouris et al. 1997; 1998;1999; Kuchinski et al. 1998; 
Ohta \& Kodaira 1995). This does not include, however, 
any stellar population with a scaleheight small enough to be effectively 
completely hidden by the dust. Such a stellar population will be discussed in
Sect. 2.2.3.  

NGC891 was modeled (Xilouris et al. 1998,1999) by fitting an artificial image 
produced by radiative transfer on observations in an attempt to conclude on 
the dust morphology. In this section we will adopt the
latest model for NGC891 (Xilouris et al. 1999). 
In this model, the emissivity is described by an exponential disk and a 
de Vaucouleurs bulge, while the dust resides in a pure exponential disk. 
The parameters describing the stars and the dust are determined in 5 
wavebands (B, V, I, J, K).

For the stellar emissivity the following formula is used
\begin{displaymath}
\eta_0(R,z) = L_s \exp \left( - \frac{R}{h_s} - \frac{|z|}{z_s} \right)
\end{displaymath}
\begin{equation}
~~~~~~~~~~+L_b \exp (-7.67\,B^{1/4})\,B^{-7/8}
\end{equation}
In this expression the first part describes an
exponential disk, and
the second part describes the bulge, which in projection is the well-known
$R^{1/4}$-law (Christensen 1990).
Here $R$ and $z$ are the 
cylindrical coordinates, $L_s$ is the stellar emissivity per unit volume per
steradian at the center of the disk and $h_s$ and $z_s$ are the 
scalelength and 
scaleheight respectively of the stars in the disk.
For the bulge, $L_b$ is a normalisation constant,
while $B$ is defined by
\begin{equation}
B = \frac{\sqrt{R^2 + z^2\,(a/b)^2}}{R_e} 
\end{equation}
with $R_e$ being the effective radius of the bulge and $a$ and $b$ being the
semi-major and semi-minor axis respectively of the bulge. In this paper we will
refer to the exponential disk described above as to the old stellar disk, in 
order to differentiate it from the young stellar disk emitting in UV 
(Sect. 2.2.3).

For the dust distribution a similar expression as that
adopted for the stellar distribution in the disk is used, namely
\begin{equation}
\kappa_{ext}(\lambda,R,z) = \kappa_{ext}(\lambda,0,0)\,\exp \left( - \frac{R}{h_d}
- \frac{|z|}{z_d} \right)
\end{equation}
where $\kappa_{ext}(\lambda,0,0)$ is the extinction coefficient at the center 
of the disk for the wavelength considered, and $h_d$ and $z_d$ are the 
scalelength and scaleheight respectively of the dust. The central optical 
depth of the model galaxy seen face-on is 
\begin{eqnarray}
\tau^f(\lambda) = 2\,\kappa_{ext}(\lambda,0,0)\,z_d
\end{eqnarray}
The parameters determined for NGC891 are presented in
Table 1. The total amount of dust derived from this model is 
$M_{dust} = 5.6\times 10^7\,{\rm M}_{\odot}$ (Xilouris et al. (1999). 

\begin{table} 
\caption{Parameters of NGC891 derived by Xilouris et al. (1999).}
\begin{tabular}{lccccccc}
\hline
Param.                                    &  B  &  V  &  I  &  J  &  K  \\ 
\hline 
$L_s$ [$\frac{erg}{sec pc^3 ster}\times10^{27}$] &2.66 &3.53 &3.44 &6.21 &1.41 \\
$z_s$ [kpc]                                     &0.43 &0.42 &0.38 &0.43 &0.34 \\         
$h_s$ [kpc]                                     &5.67 &5.48 &4.93 &3.86 &3.87 \\          
$L_b$ [$\frac{erg}{sec pc^3 ster}\times10^{30}$] &12.0 &7.4  &2.23 &4.99 &1.71 \\
$R_e$ [kpc]                                     &1.12 &1.51 &1.97 &0.87 &0.86 \\
$b/a$          --                               &0.60 &0.54 &0.54 &0.71 &0.76 \\
$\tau^f$   \,\,\,--                                & 0.87&0.79&0.58&0.23&0.10\\
$z_d$          [kpc]                            &0.27 &0.27 &0.27 &0.27 &0.27 \\        
$h_d$          [kpc]                            &7.97 &7.97 &7.97 &7.97 &7.97 \\   
\hline
\end{tabular}
\end{table}

The extinction coefficients $\kappa_{ext}(\lambda,0,0)$ 
were derived as fitting parameters 
(Xilouris et al. 1999), separately for each of the five wavebands
used for
modelling the galaxy. They were then used to calculate the absorption of the
optical light by dust, via the radiation transfer procedure. Since in 
calculating the emission of the dust we used the absorption efficiencies 
($Q_{abs}$) taken from Laor \& Draine (1993), as presented in Sect. 2.1, we
compare in Fig. 1 the extinction coefficients derived from the 
fitting routine with the theoretical ones derived from the extinction 
efficiencies (Eqs. [2] and [3]). The result of such a comparison shows that
indeed the extinction derived from observations (square symbols) is very 
close to the theoretical one used to calculate the IR emission 
(solid line). This ensures that the treatment of the absorption and of the
emission is consistent. In Fig. 1 we also give the wavelength dependence
of the absorption and scattering coefficients, plotted with dashed and dotted
lines, respectively. 

\begin{figure}[htp]
\includegraphics[scale=0.50]{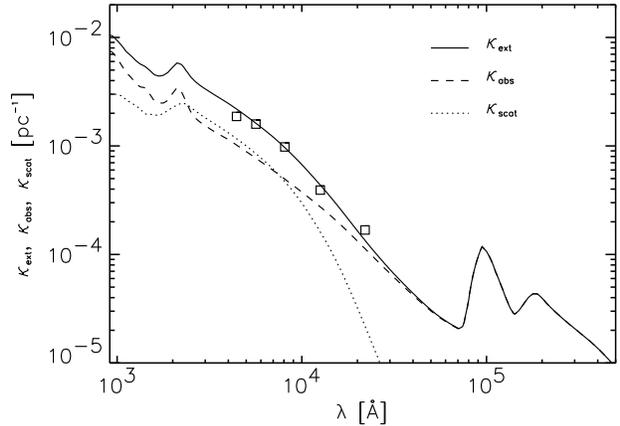}
\caption{The wavelength dependence of the extinction (solid line), 
absorption (dashed line) and scattering (dotted line) coefficients derived 
from the extinction, absorption and scattering efficiencies taken from Laor 
\& Draine (1993). The symbols represent the extinction coefficients derived as
fitting parameters by Xilouris et al. (1999).}
\end{figure}

\subsubsection{UV radiation field} 

The UV radiation field cannot be derived directly from  observations. For 
edge-on galaxies it is extremely difficult to infer something about the young 
stellar disk, which, due to it's small scaleheight, is presumably highly 
obscured by line-of sight dust. Any young stellar population whose 
scaleheight 
is smaller than the scaleheight of the diffuse dust will be totally 
unprobed by UV, optical and NIR data on an edge-on galaxy. Thus for an edge-on 
galaxy mid to far-IR emission would be the primary observational signature for 
any young massive star population having a scaleheight of 100\,pc or less. 
This stellar population would (we assume luminosity is  dominated by stars) 
produce almost all the non-ionising and ionising UV from the galactic disk. 

We thus consider the UV radiation field as a parameter and we 
parameterise the UV luminosity in terms of a recent SFR, based on the 
population synthesis models of Mateu \& Bruzual (2000). We consider
$Z={\rm Z}_{\odot}$, a Salpeter IMF, $M_{up}=100\,{\rm M}_{\odot}$ and 
${\tau} = 5$\,Gyr. This gives, for example, SFR $=8.55\times 10^{-28}\,L_{UV}$
[erg/s/Hz] for ${\lambda}=912\,$\AA, SFR $=3.74\times 10^{-43}\,L_{UV}$ 
[erg/s] for the integrated ionising UV shortwards of 912\,\AA, 
SFR $=1.18\times10^{-28}\,L_{UV}$ [erg/s/Hz] at $\lambda=1500$\,\AA, etc.
The diffuse UV radiation derived from the calibration above was confined in a 
thin disk, with a scaleheight 
$z=90$\,pc (close to that of the Milky Way, Mihalas \& Binney 1981) and 
the same scalelength as the blue disk. 

\begin{figure*}[htp]
\plotfiddle{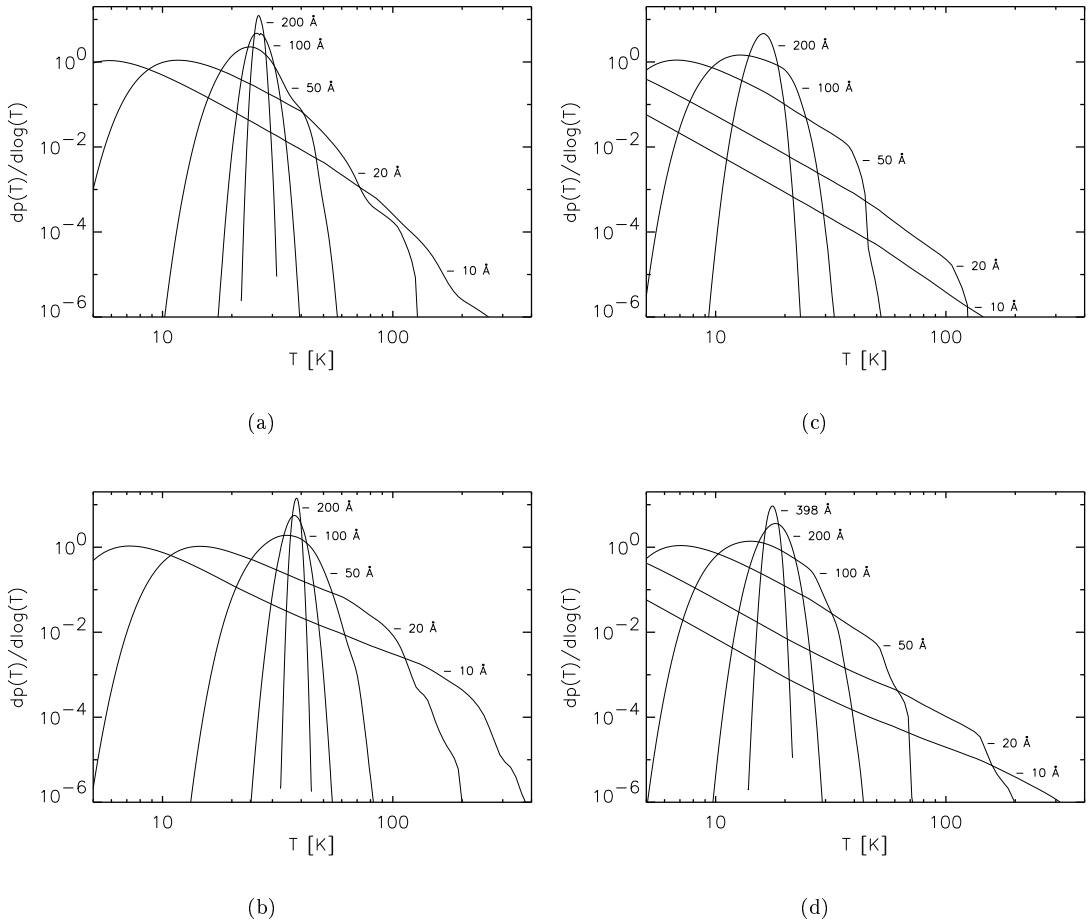}{5.in}{0.}{100.}{100.}{-350.}{-340.}
\caption[]{Examples of temperature distribution for dust grain sizes 
(10\,\AA, 20\,\AA, 50\,\AA, ...) embedded in the diffuse radiation field, for a solution with the UV 
radiation field in the diffuse component corresponding to a 
SFR $=3\,{\rm M}_{\odot}$/yr: a) silicate grains in the center of the galaxy
($R=0$\,kpc, $z=0$\,kpc); b) graphite grains in the center of the galaxy;
c) silicate grains at the edge of the galactic disk 
($R=15$\,kpc, $z=0$\,kpc); d) graphite grains at the edge of the galactic 
disk.}
\end{figure*} 

The UV luminosity could, in principle, also be indirectly 
inferred from different indicators 
of the SFR, like the free-free emission or H${\alpha}$ emission. There are 
however large uncertainties in deriving these quantities, especially in the
case of NGC891. The first indicator, the radio free-free emission, has the 
advantage of being optically thin. Niklas et al. (1997) found that the total 
radio emission of NGC891 is dominated by synchrotron emission, while the 
free-free emission constitutes a thermal fraction of just 0.05 at an 
observing frequency of 10 GHz. However, the radio morphology of 
NGC891 is rather complicated, with more morphological components having 
different spectral indices. Allen et al. (1978) suggested the existence of 
two components for the radio continuum emission: a highly flattened thin disk 
component coinciding with the equatorial plane of the galaxy and a thick disk 
or halo component (see also Hummel et al. 1991). If this is the case, then
the free-free emission estimated from a spectral decomposition of
the integrated fluxes of the galaxy (as done by Niklas et al. 1997) could 
have been seriously underestimated. We conclude that the free-free emission 
does not constitute an accurate constraint of our model.

The second indicator of the SFR is the H$\alpha$ emission. It is known that
correcting H$\alpha$ data for internal extinction is a very uncertain step,
since the extinction contains a large-scale component plus 
extinction due to localised sources, which are difficult to account for in 
edge-on disks. Hoopes et al. (1999) found that in NGC891, $83 - 86\%$ of the
observed H$\alpha$ emission is in the diffuse component associated with the
halo of the galaxy. This would imply that most of the emission from the disk is
highly obscured, making unreliable any derivation of the SFR in the disk.
 
\subsection{Calculation of the IR spectrum from grains embedded in the
diffuse optical and UV radiation field}

\begin{figure}[htb]
\includegraphics[scale=0.45]{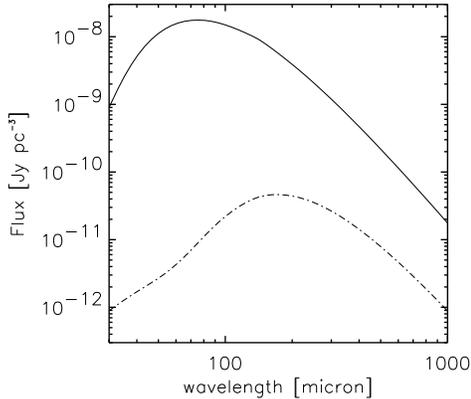}
\caption[]{The infrared spectrum from dust embedded in the diffuse 
radiation field, for a solution with the UV radiation field in the diffuse 
component corresponding to a SFR $=3\,{\rm M}_{\odot}$/yr. The figure
illustrates the infrared spectrum obtained for two extreme locations in the
galaxy:  in the center of the galaxy ($R=0$\,kpc, $z=0$\,kpc) - plotted with
solid line and at the edge of the galactic disk ($R=15$\,kpc, $z=0$\,kpc) -
plotted with dashed-dotted line.} 
\end{figure} 

The dust heating due to photons from the diffuse interstellar radiation field 
was calculated following the method of Guhathakurta \& Draine (1989). This 
method derives the temperature distribution $P(a,T)$ of various grain 
radii $a$ as a function of dust temperature $T$. Due to stochastic heating, 
small grains undergo significant fluctuations from the equilibrium 
temperature, 
while larger grains have narrower probability functions, eventually 
approaching delta functions. We calculated the temperature
distribution at any point in the galaxy, based on the energy density of the
radiation field derived in the previous subsections.  The 
spectral shape of the UV-NIR energy density and its magnitude will 
obviously determine the shape of the
temperature distributions. The grain sizes used in the calculations were chosen
between $a_{min}=10\,$\AA\, and $a_{max}=0.25\,{\mu}$m, for a step
${\Delta}{\rm log}\,a=0.05$. In Fig. 2 we give the temperature distribution as 
${\rm d}p/{\rm dlog}T$ for some silicate (Fig. 2a,c) and graphite 
(Fig. 2b,d) grain sizes   
embedded in the diffuse radiation field, for a solution with the UV 
radiation field in the diffuse component corresponding to a 
SFR $=3\,{\rm M}_{\odot}$/yr. Fig. 2  also illustrates the dependence of the 
grain
temperature on the position in the disk. We give two examples, 
for two extreme positions in the plane of the disk: in the center of the disk 
(Fig. 2a,b) and at the edge of the disk (Fig. 2c,d). At the edge of the disk 
the probability distribution converges
towards lower equilibrium temperatures, as expected due to the decrease in the 
intensity of the radiation field. 
We also note the relatively large difference in the equilibrium
temperature between $\sim30$\,K for the central
region and $\sim15$\,K for the outer
disk. This is interesting in the context of the results of Alton et al. (1998)
who fitted the SED of NGC891 with 15 and 30\,K grey-body components. We discuss
this further in Sect.\,7, in terms of the geometry and opacity
of the disk.
We also note that in passing from the centre to the
outer regions, progressively 
bigger grains exhibit stochastic heating, as expected due to the lower energy
density of the radiation field. For the same location in the
disk, graphite grains undergo larger fluctuations than the silicate grains.
 
Given the temperature probability function $P(a,T)$, the IR flux is
given by

\begin{eqnarray}
F_{\nu} & = & \frac{1} {d^2}\int_{a_{\rm min}}^{a_{\rm max}} 
N(a)\,{\rm d}a\,{\pi}\,a^2\,Q_{\rm abs}{(\nu},a)
\nonumber \\
& & \times\int_0^{\infty}B_{\nu}(T)\,P(a,T)\,{\rm d}T  
\end{eqnarray}
\\
where $B_{\nu}$ is the Planck function, $N(a)$ is the grain size distribution
given by Eq. (1), and $d$ is the distance to the galaxy.

The IR emission was calculated in each point of the galaxy and then the
total emission was obtained by integrating the emission over the galaxy. In
Fig. 3 we give two examples of IR spectra for two extreme locations in
the plane of the galactic disk, namely in the center of the galaxy 
($R=0$\,kpc, $z=0$\,kpc) and at the edge of the disk 
($R=15$\,kpc, $z=0$\,kpc). At the edge of the disk the infrared 
emission is colder, as expected due to the decrease in the dust temperature, 
though the increased importance of stochastic heating in the outer regions 
tends to flatten the SED on the Wien side.

\section{Results for the \lq\lq standard model\rq\rq}

The FIR emission for our \lq\lq standard model\rq\rq\, was 
calculated for a grid of SFR, and the different solutions were compared with 
the observations. To extract the FIR luminosity of the galactic dust we 
integrated the SED of NGC891 between 40 and 1000$\,{\mu}$m, with the data 
taken from Alton et al. (1998). The SED was 
parameterised by Alton et al. (1998) by a
fitting two-temperature greybody curve (with a wavelength dependence of grain 
emissivity in the FIR $\beta=2$) to the observed fluxes 
from IRAS and SCUBA at 60, 100, 450 and 850\,${\mu}$m. This yielded 
best-fit ``temperatures'' of  
30 and 15\,K to the observed spectrum, and an integrated luminosity between 
40 and 1000$\,{\mu}$m of $5.81\times 10^{36}$\,W. This output luminosity 
can be
accommodated in our ``standard model'' with a solution for a 
SFR $=7.5\,{\rm M}_{\odot}$/yr. The predicted FIR SED from the model
is given in Fig. 4a, together with the UV-optical-NIR intrinsic emitted stellar
radiation (as would have been observed in the absence of dust). The IRAS and SCUBA data (diamonds) from  Alton et al. (1998) are
also given for comparison. The inferred SFR is an upper limit, since we assumed
that all the non-ionising UV escapes into the diffuse component. Indeed, in the 
absence of local absorption, all the lines of sight from each source
would contribute primary photons to the diffuse interstellar radiation field.
In practice, some fraction of the lines of sight - call this fraction 
$F$ - from each source will be opaque to local absorption, giving rise to 
discrete IR sources embedded in a diffuse IR emitting disk. 
Our solution was
thus calculated  under the extreme assumption that $F=0$ for the non-ionising 
UV and $F=1$ for the ionising UV. While requiring a solution with a large
SFR, the predicted SED is slightly warmer than the observed
SED at 100\,${\mu}$m. However, the main difficulty of the model is that it  
underestimates by a factor of 2 to 4 the observed sub-mm fluxes.

\section{Results for the \lq\lq standard model\rq\rq\, supplemented by
localised sources}

\begin{figure*}[htp]
\subfigure[] {
\includegraphics[scale=0.45]{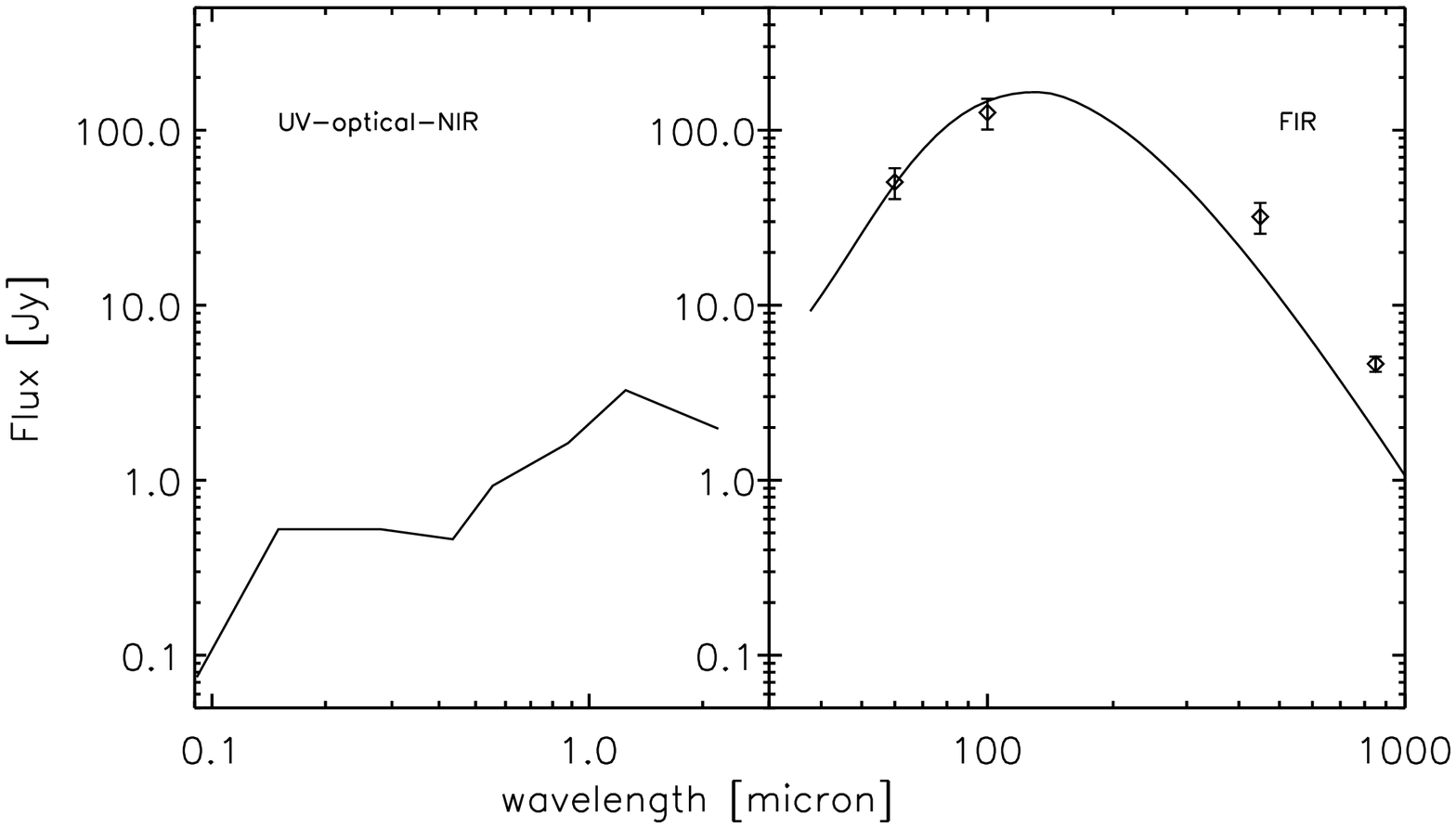}}
\subfigure[] {
\includegraphics[scale=0.45]{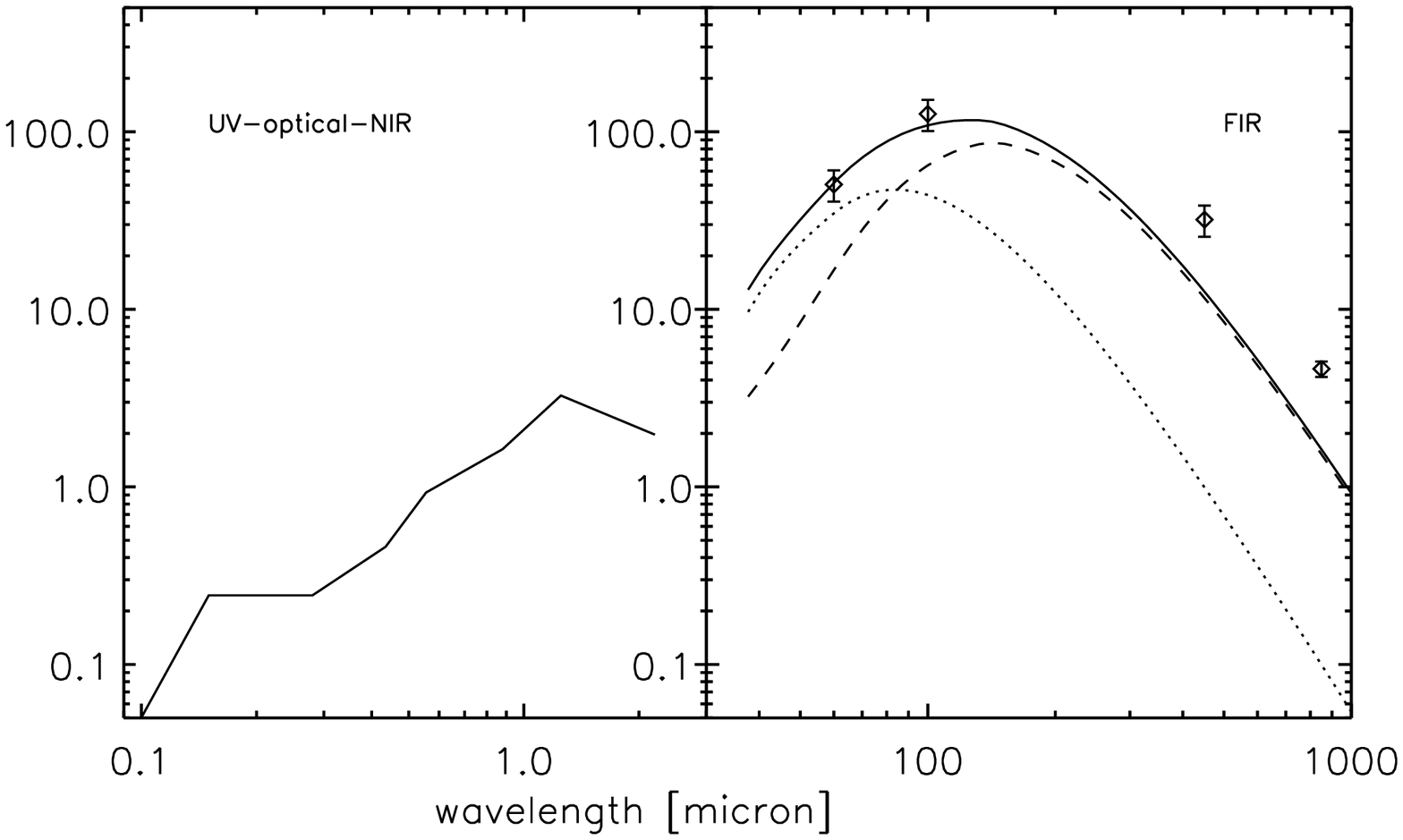}}
\caption{a) The SED from our \lq\lq standard model\rq\rq, for a current 
SFR $=7.5\,{\rm M}_{\odot}$/yr. The UV-optical-NIR SED plotted in the left hand
panel represents the intrinsic emitted stellar radiation (as would have been
observed in the absence of dust) while the FIR-sub-mm
curve plotted in the right hand panel is the predicted SED for the re-radiated 
dust emission. The SFR has been adjusted such that
the predicted FIR model luminosity equals the observed one. 
The observed FIR fluxes from Alton et al. (1998) are given as diamonds.
b) The SED from our \lq\lq standard model\rq\rq\, supplemented by localised 
sources, for a SFR $=3.5\,{\rm M}_{\odot}$/yr and $F=0.28$.  
The UV-optical-NIR SED plotted in the left hand
panel represents the intrinsic emitted stellar radiation while the FIR-sub-mm
curve (solid line) plotted in the right hand panel is the predicted SED for 
the re-radiated dust emission. In the right hand panel we also show the
contribution of the diffuse component (dashed-line) and of the HII
component (dotted line) to the total predicted FIR emission. The observed 
FIR fluxes from Alton et al. (1998) are again given as diamonds.}
\end{figure*} 

A more realistic solution is to consider that some fraction $F$ of the 
non-ionising UV is locally absorbed in star-forming complexes 
(e.g., HII regions). Such a solution requires two free parameters ($F$ and 
SFR) and can be accommodated with a lower SFR, due to the higher
probability of absorption of the non-ionising UV photons.  To include the
contribution of HII regions we used a template SED of such forming complexes.
We utilised the galactic UC HII region G45.12+0.13, with FIR observations at
60 and  100\,${\mu}$m from the IRAS Point Source Catalog and 1300\,${\mu}$m
measurements from Chini et al. (1986).  We fitted the observed data 
with a greybody curve (with a wavelength dependence of grain emissivity in 
the FIR ${\beta}=2$) and used it as a template SED for the compact source
component for our modelling. 
We do not attempt here to include in this template potential cold emission
components that might be expected from ``parent'' molecular clouds in 
juxstaposition to their ``offspring'' HII regions.

The best solution was obtained for a SFR $=3.5\,{\rm M}_{\odot}$/yr and 
$F=0.28$. The 
predicted FIR SED is given in Fig. 4b, again together with the UV-optical-NIR
intrinsic stellar radiation. In Fig. 4b we also give the contribution of the 
diffuse component (dashed line) and of the HII 
component (dotted line) to the the total predicted FIR SED (solid line).  The 
integrated emitted luminosity
predicted by the model is $L=4.79\times 10^{36}$\,W, which accounts for 
$82\%$ of the total observed emitted FIR luminosity. Of this, the diffuse 
component
contributes $46\%$ of the total observed dust luminosity and the
localised sources $36\%$. However, the main difficulty of the model is 
again that the solution fails to reproduce the observations longwards of 
100\,${\mu}$m.

\section{The missing FIR/sub-mm component of our \lq\lq standard model\rq\rq}

We have shown that our \lq\lq standard model\rq\rq\, fails to
reproduce the observed SED of NGC891 in the sub-mm spectral range.
The more realistic case, where the contribution from HII regions is added to the
diffuse component, is also not able to predict the observed FIR/sub-mm SED. 
However, this may be, to some extent, because we did not include a cold dust 
emission component that might be expected to arise from associated 
parent molecular clouds. 

Below we consider four more ways that might account for this 
discrepancy. Firstly, we discuss the effect of altering the 
geometry of the large-scale distribution of stars and dust. Alternatively, we 
consider extraplanar emission. Thirdly, the predicted apparent sub-mm/optical 
ratio could be increased by including more grains in the disk, but 
distributed in clumps, so as not to effect the optical extinction properties 
of the diffuse disk. Lastly, a more radical solution is presented, namely a 
model with two dust-disk components, which is a 
moderately optically thick solution.

\subsection{The effect of altering the geometry of the large-scale 
distribution of stars and dust}

The diffuse IR emission predicted by our model was derived for an
exponential disk of stars and dust, adopted from Xilouris et al. (1999). 
Galactic disks are however known to be quite complex systems, where the 
large-scale 
distribution of stars and dust presents inhomogeneities in the form of spiral
structures. Adding logarithmic spiral arms as a perturbation on the 
exponential disks for stars and dust would alter the solutions of the 
radiative transfer, and thus of the calculated diffuse IR emission. 
However, Misiriotis et al. (2000b) showed that plain
exponential disk models give a very accurate description for galactic disks
seen edge-on, with only small deviations in parameter values from the real ones
(typically a few percent). Thus, we conclude that the geometry adopted here
cannot be responsible to account for the discrepancy between our 
model predictions and the observations.

\subsection{Extraplanar dust?}

Observations of external galaxies have revealed extensive thickened layers of
ionised gas traced by its H$\alpha$ emission in several edge-on spiral 
galaxies (e.g., Rand et al. 1990, 1992; Dettmar 1990; Pildis et al. 1994; 
Rand 1996, Ferguson et al. 1996; Hoopes et al. 1999). 
Amongst the most spectacular examples of
extraplanar diffuse ionising gas (DIG) is that seen in NGC891. 
The recent observations of Hoopes et
al. (1999) showed that the DIG layer in NGC891 extends out to at least 5 kpc
from the plane, and possibly as far as 7 kpc, and has the brightest and largest
DIG layer known. 

In addition to detecting the high z DIG studied in earlier work, Howk \& 
Savage (2000) also detected individual dust-bearing clouds observable to 
heights $z\sim 2$\,kpc from the midplane. 
Moreover, they detected the presence of HII regions at large 
distances from the midplane (0.6-2\,kpc), which suggests that on-going star 
formation may be present in some of the dense, high-z clouds. 

In principle both the dust-bearing clouds ($z\sim2\,$kpc) seen in absorption 
as well as the dust that may be associated with the DIG ($z\sim 5\,$kpc) 
could contribute to the
FIR/sub-mm emission of the galaxy. The possibility of the extraplanar 
dust
clouds emitting in the sub-mm was recently investigated by Alton et
al. (2000b). From the upper limits in the total amount of extraplanar
dust these authors found that less than $5\%$ of galactic dust exists 
outside the 
galactic disk, if the dust grains are not colder than 17\,K, or $9\%$, if the 
dust temperature is 10\,K. This small percentage cannot account for 
our discrepancy between the model and the observations. 

We also consider it unlikely that any diffuse dust component 
associated with the DIG ($z\sim 5\,$kpc) could provide a 
substantial fraction of the sub-mm emission.
In that case, if the extraplanar emission were powered by the absorption of 
disk photons, the optical depth of the DIG would have to be comparable to 
that of the disk, which would be a very extreme scenario, probably
also in conflict with the constraints on submillimeter extraplanar emission.

\subsection{The case of hidden dust in clumps}

One possible explanation to account for the missing dust component
could be that Xilouris et al. (1999) underestimated the dust content
of the galaxy because of the adopted diffuse distribution 
for the dust. Since the galaxy is optically thin, any additional
dust component would have to be distributed in very dense clumps 
instead of being diffuse. Only then would a significant fraction of the 
lines of sight avoid the additional dust, so as not to affect 
the derived optical-NIR optical depth. 

Such dust clumps could either be a component of star-formation regions, 
or could have no associated sources - we refer to the latter hypothesis as the
\lq\lq quiescent clumps\rq\rq. 
The quiescent clumps must be optically thick to the 
diffuse UV/optical radiation field in the disk to have an impact on
the predicted submillimeter emission.
They would radiate at predominantly longer 
wavelengths than the diffuse disk emission, in the FIR/sub-mm spectral range.
One can speculate that such optically thick ``quiescent clouds'' 
could be physically identified with partially or wholly collapsed clouds 
that, for lack of a trigger, have not (yet) begun to form stars. However,
due to the 
lack of intrinsic sources, a very substantial mass of dust would have to be 
associated with the quiescent clouds to account for the submillimeter 
emission missing from our model prediction.

In
reality there must be also dark clouds associated with star-forming complexes.
In the Milky Way HII regions around newly born massive stars are commonly 
seen in juxtaposition to parent molecular clouds (e.g., M17). This is thought 
to be a consequence of the fragmentation of the clouds due to mechanical 
energy input from the winds of the massive stars. Thus, warm dust emission from
cloud surfaces directly illuminated by massive stars can be seen along
a fraction of the lines of sight, together with cold sub-mm
dust emission from the interior of the associated optically thick cloud 
fragments. These dense clumps (with or without associated sources) may 
contribute to the sub-mm emission, and thus supplement the contribution of
compact HII regions.

However, the question remains whether it is possible to place
such a substantial amount of additional dust in clumps without affecting the 
opacity of the galaxy. Kuhinski et al. (1998) modeled a sample of highly 
inclined galaxies using both smooth and clumpy dust and concluded that
clumpiness does not effect dramatically the opacity of highly inclined
galaxies. On the other hand, Bianchi et al. (1999) adopt different
parameterisation for clumps and they report that clumpiness does
significantly effect the opacity of edge-on galaxies. It is beyond the scope of
this paper to calculate a quantitative solution for a clumpy distribution, 
but we qualitatively consider the clumpy scenario, as quiescent clumps
or associated with star-formation regions - to be a possibility to 
account for the missing FIR-sub-mm component.

\subsection{A two-dust-disk model}

We have seen that the observed sub-mm SED requires  
more dust in the galaxy than predicted by our ``standard model''.
We have already discussed the possibility of including
this extra dust in clumps. An alternative solution is obtained if it is 
postulated that this additional dust is confined in a diffuse thin disk, 
in which the young stellar population is embedded. 
(A thin young stellar disk, but with no associated dust, was considered in
Xilouris et al. 1998.)  If this second dust-disk 
component has a scaleheight comparable with the 90\,pc scaleheight
of molecular gas clouds in the Milky Way, it would not have been 
seen in the optical/NIR analysis of Xilouris et al. (1999), being
totally obscured by the optically visible
disk of dust. Indeed, since the galaxy is almost 
edge-on, adding a second thin disk of dust will probably
not change the parameters 
for the intrinsic distributions of stars and dust derived by Xilouris et al.
(1999).  However, adding a second disk of dust
will change the optical depth of the disk, transforming our solution into a
moderately optically thick solution ($\tau^f_V=3.1$).

\begin{figure}[htp]
\includegraphics[scale=0.45]{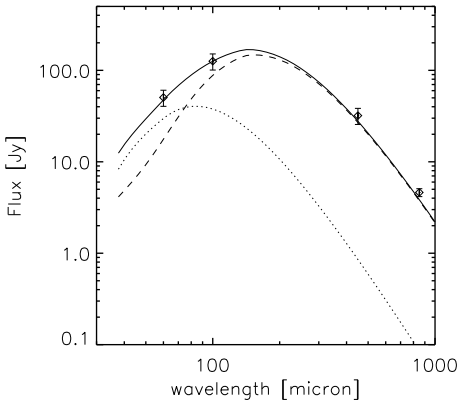}
\caption{The predicted FIR emission from a two-dust-disk model, for a 
SFR $=3.8\,{\rm M}_{\odot}$/yr and $F=0.22$. The total predicted SED is given 
with solid line, the
diffuse component with dashed line and the HII component with dotted line. 
The observed fluxes from Alton et al. (1998) are given 
as diamonds.}
\end{figure} 

We have calculated the energy density of the radiation field from UV to 
NIR for such a two-dust-disk model, and subsequently the infrared
emission. The best solution was obtained for a SFR $=3.8\,{\rm M}_{\odot}$/yr,
$F=0.22$ and $M_{dust}=7\times 10^{7}\,{\rm M}_{\odot}$ 
in the second disk of dust. The corresponding non-ionising UV  luminosity for
this SFR is $\sim 8.2\times 10^{36}$\,W. This could be provided, for example, 
by a population of B5 stars ($T_{\rm eff}=15500$\,K) with a space density in 
the centre of the galaxy of $7.5\times 10^{-4}\, {\rm pc}^{-3}$.
The predicted combined SED from the two diffuse dust-disk 
components is shown by the dashed line in Fig. 5. By adding the further  
contribution from localised sources (dotted line), as discussed in Sect. 4, 
we were able to fit the SED of NGC891, from the FIR to the sub-mm. The 
luminosity of the diffuse component is $4.07\times 10^{36}$\,W, which 
accounts for $69\%$ of the observed FIR luminosity, and the luminosity of the 
HII component is $1.82\times 10^{36}$\,W, making up the remaining $31\%$ 
of the FIR luminosity. This model can thus successfully fit the shape of the
SED. In the next section we will show that the predictions of the 
two-dust-disk model are in agreement with the observed radial profile at 
850\,${\mu}$m. Nevertheless, further modelling of face-on galaxies will be
required to test the validity of the two-dust-disk model. Ultimately, the 
two-dust-disk hypothesis will be directly tested observationally using the 
new generation of sub-mm interferometers, which will resolve edge-on disks.

\section{The radial profiles}

\begin{figure}[htb]
\includegraphics[scale=0.45]{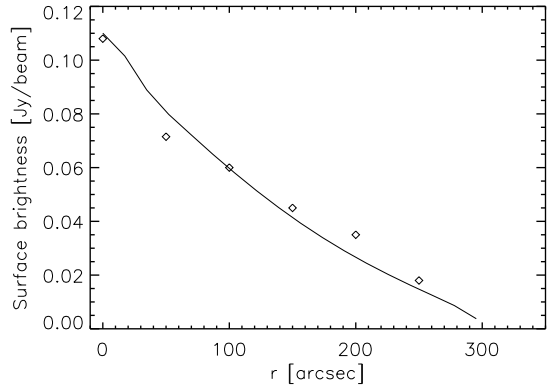} 
\caption[]{The averaged radial profile at $850\,\mu$m from the two-dust-disk model, plotted
with the solid line. The profile is averaged over a bin width of $36^
{\prime\prime}$, for a sampling of $3^{\prime\prime}$ and for a beam width of
$16^{\prime\prime}$, in the same way as the
observed averaged radial profile from Alton et al. (2000a) (plotted with
diamonds).} 
\end{figure}

\begin{figure*}[htb]
\plotfiddle{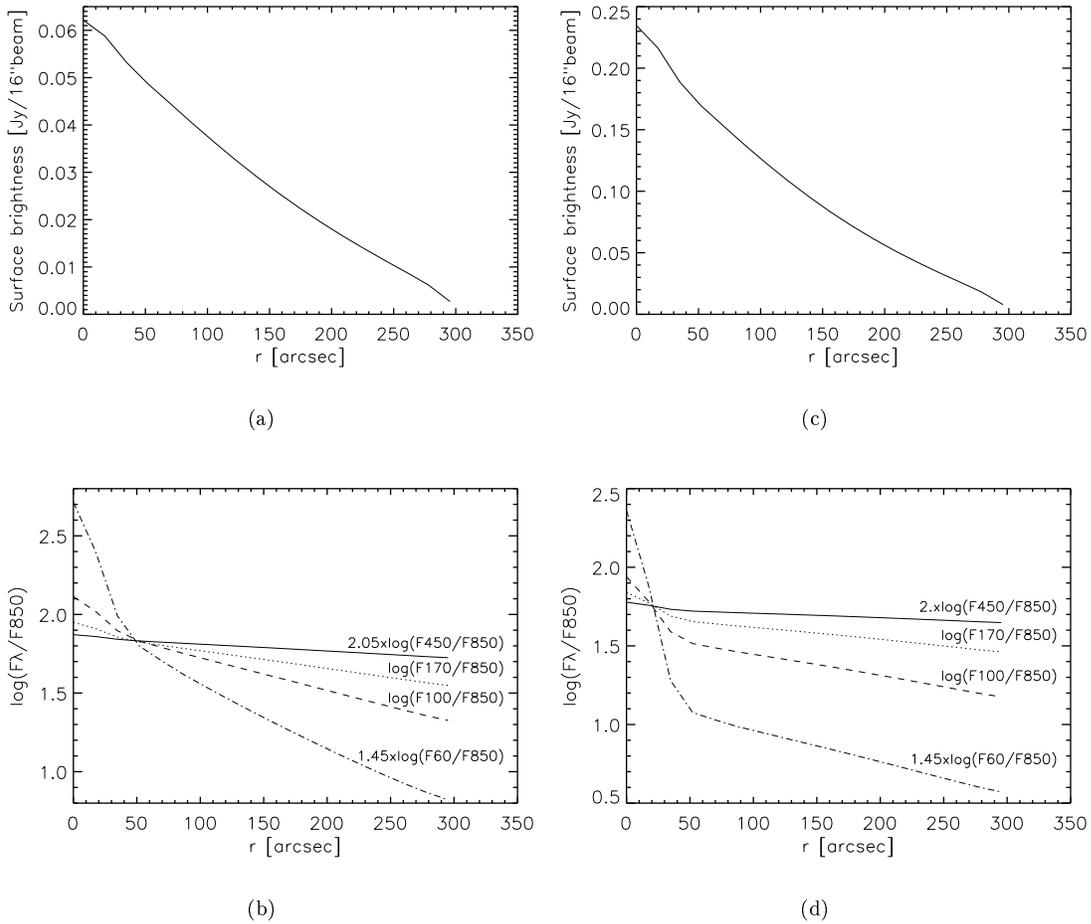}{5.0 in}{0.}{100.}{100.}{-350.}{-340.}
\caption[]{a) The predicted diffuse FIR radial profile from our \lq\lq standard
model\rq\rq\, at 850 $\mu$m, for SFR $=3.5\,{\rm M}_{\odot}$/yr. The radial 
distance
is given in arcsec, with 100$^{\prime\prime}\sim5$\,kpc). b) The predicted 
colour profiles  at different wavelengths, for the same model like in a). 
The ratios were multiplied by the given factors to provide the same colours 
at an angular radius of 50$^{\prime\prime}$. The colour profiles show that 
for shorter wavelengths the effect of decreasing the dust temperature towards 
larger radii become more and more pronounced. The colour profile F60/F850 has 
the steepest gradient, with a very hot component within the inner 50 arcsec 
radius, followed by the increase of the cold dust at the larger 
galactocentric distances. c) The predicted diffuse FIR radial profile from 
our two-dust-disk model, at 850 $\mu$m, for SFR $=3.8\,{\rm M}_{\odot}$/yr. 
d) The
colour profiles for the two-dust-disk model. Again, the ratios were 
multiplied by the given factors to provide the same colours at an angular 
radius of 20$^{\prime\prime}$.    
} 
\end{figure*}

Since we have calculated the 3 dimensional FIR radiation field in NGC891, for
both our \lq\lq standard model\rq\rq\, and the two-dust-disk model, it is
straightforward to integrate along the line of sight and thus produce a 
2-dimensional map of the galaxy at different wavelengths, and then radial 
profiles. Here we consider only the diffuse component of the IR 
emission. Because observed radial profiles were derived by Alton et al. 
(2000a) using the SCUBA observations at $850\,\mu$m, we first attempt to
calculate the radial profiles at this wavelength and compare it with the
observations. The calculations were done at high resolution and afterwards 
smoothed to 16$^{\prime\prime}$ resolution.
 Since the observed radial profile at $850\,\mu$m from 
Alton et al. (2000a) were derived as averages over a bin width of
36$^{\prime\prime}$, and with a sampling of 3$^{\prime\prime}$, we have also
calculated an averaged radial profile in the same way as Alton et al. (2000a)
did, and directly compare it with the observations. We have found that in the
case of the two dust disk model there is a very good agreement between the
model predictions and the observations and this comparison is illustrated in 
Fig. 6. The predicted radial profile can 
be traced out to 300 arcsec radius (15 kpc), as also detected by the SCUBA.

In Figs. 7a and 7c 
we give our prediction for the radial profile of NGC891 
at 850 $\mu$m, this time as it would be observed within a beam of
16$^{\prime\prime}$, and without any averaging. This is shown for 
our \lq\lq standard model\rq\rq\, and the two-dust-disk model, respectively. 
In Fig. 7b.d we give the radial colour profiles at the wavelengths used by the 
IRAS, ISO and SCUBA, normalised to the radial profile at 850 $\mu$m. The 
strong colour gradients, seen in particular between the 60 or 100 $\mu$m
and the sub-mm regime, are a consequence of the strong gradients in the
radiation field calculated in our finite disk model. It is interesting to note
that the stronger contribution of stochastic emission from small grains
(relative to big grains) in the outer disk at 60 $\mu$m, does not strongly 
influence the gradients. Potentially, therefore, large-scale colour 
gradients in galaxy disks, could act as a diagnostic for the relative 
contribution of the diffuse emission and compact sources, as the latter would
not be expected to have IR spectra strongly dependent on position in the disk. 
As expected, the 450/850 colour shows only a shallow gradient, as both these
wavelengths are displaced well longwards of the spectral peak, and thus depend 
only approximately linearly on grain temperature.

\section{Discussion: the contribution of different stellar populations in 
heating the dust in NGC891}

\begin{figure*}[htp]
\plotfiddle{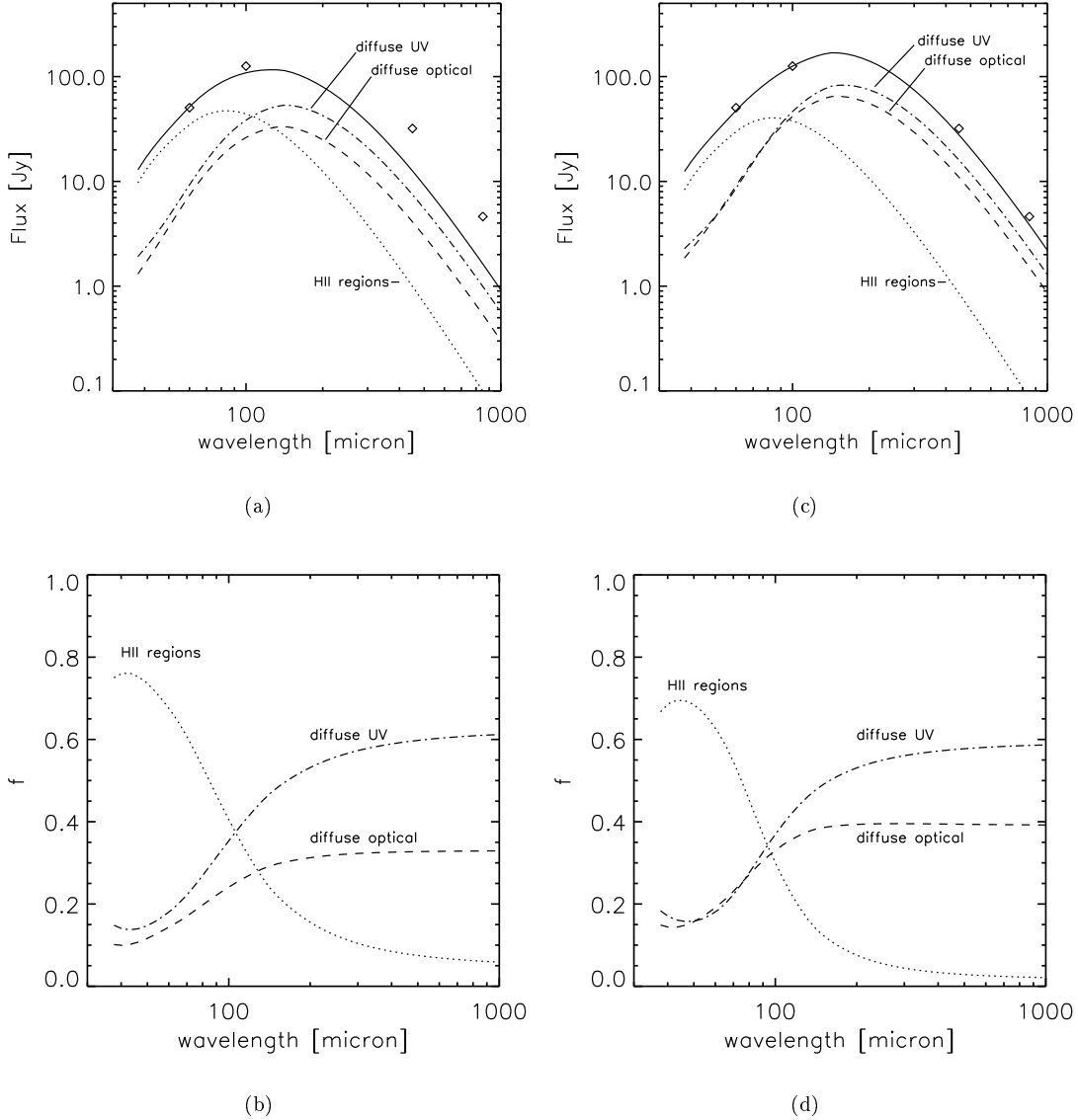}{6.in}{0.}{100.}{100.}{-350.}{-230.}
\caption[]{a) The contribution of different stellar components to the FIR 
emission, as a function of wavelength, for our \lq\lq standard model\rq\rq. The
legend is as follows: dashed-line: contribution of diffuse optical radiation 
($4000-22000$\,\AA), dashed-dotted line: contribution of diffuse UV 
radiation ($912-4000$\,\AA), dotted-line: contribution of HII component, 
solid line: total predicted FIR SED. The observed fluxes from Alton et al. 
(1998) are given as diamonds.
b) The same as in case a), but here the different
contributions are given as the fraction f of the FIR emission produced by a
certain component to the total FIR predicted emission. c) and d) are the same
as a) and b), respectively, but for the two-dust-disk model.} 
\end{figure*} 

The relative contribution of optical and UV photons in heating the dust has 
been a long
standing question in the literature. Since we have a detailed calculation of 
the absorbed energy over the whole spectral range and at each position in the
galaxy, we can directly calculate which part of the
emitted FIR luminosity from each volume element of the galaxy is due to
the optical and NIR
photons, and which part is due to the UV photons. The
IR emission from any given volume element arises from a common local grain 
population, so that the IR colours from that volume element will be the same 
for the emitted IR photons regardless of whether the absorbed energy 
had an optical or UV origin. However the relative fraction of energy 
absorbed from the UV and optical light varies with position in the galaxy, 
as does the dust temperature, so that the volume-integrated IR spectrum 
representing re-radiated optical light will differ from that of the 
re-radiated UV light. In this way, some information on the stellar
populations heating the grains is preserved in the IR domain, in the form
of the observed IR colours.
Volume-integrated IR 
spectral components arising from re-radiated optical and UV light are presented
in Fig. 8a,b for the \lq\lq standard model\rq\rq\, and in Fig. 8c,d for the 
two-dust-disk model, respectively. The overall contribution of
the diffuse optical and NIR
radiation ($4000-22000$\,\AA) to the total FIR luminosity
is $1.07\times10^{36}$\,W, or $22\%$ for the \lq\lq standard model\rq\rq. 
Since the predicted luminosity of this model is lower than the
observed one, it is also meaningful to give the contribution of the diffuse 
optical radiation to the observed FIR luminosity, which is $18\%$. This 
component of the FIR SED 
is plotted with dashed line in Fig. 8a,b. The
corresponding optically generated IR luminosity for the two-dust-disk model is 
$1.85\times10^{36}$\,W, or $31\%$, higher than in the \lq\lq standard 
model\rq\rq, largely because the galaxy is becoming optically thick in the 
optical range. 
This component of the FIR SED is again plotted with dashed-line in
Fig. 8c,d. The contribution of different
stellar components to the total FIR luminosity for the two dust models is 
summarised in Table 2. Table 3 gives the contribution of different stellar
components to the FIR wavelength commonly used by IRAS, ISO and SCUBA.

\begin{table}

\caption{The contribution of the diffuse optical/NIR radiation $L^{opt}_{FIR}$ 
($4000-22000$\,\AA),
 diffuse UV radiation $L^{UV}_{FIR}$ ($912-4000$\,\AA) and UV photons locally 
absorbed in HII regions $L^{HII}_{FIR}$ to the total predicted FIR luminosity 
$L^{model}_{FIR}$.}

\vspace{0.5cm}

\begin{tabular}{|l|l|l|}
\hline\hline
& & \\
 & standard model & two-dust-disk model \\
 & ${\rm SFR}=3.5\,{\rm M}_{\odot}$/yr  & ${\rm SFR}=3.8\,{\rm M}_{\odot}$/yr\\
 & $F=0.28$ & $F=0.22$\\
\hline
& & \\
$L^{opt}_{FIR}$  & $1.07\times 10^{36}$\,W & $1.85\times 10^{36}$\,W\\
                 & $22\%\,L^{model}_{FIR}$ &$31\%\,L^{model}_{FIR}$\\
\hline
& &\\
$L^{UV}_{FIR}$   & $1.60\times 10^{36}$\,W & $2.22\times 10^{36}$\,W\\
                 & $33\%\,L^{model}_{FIR}$ &$38\%\,L^{model}_{FIR}$\\
\hline
& &\\
$L^{HII}_{FIR}$  & $2.12\times 10^{36}$\,W & $1.82\times 10^{36}$\,W\\
                 & $44\%\,L^{model}_{FIR}$ &$31\%\,L^{model}_{FIR}$\\
\hline
& &\\
$L^{model}_{FIR}$   & $4.79\times 10^{36}$\,W & $5.89\times 10^{36}$\,W\\
\hline
&  &\\
$L^{observed}_{FIR}$ & $5.81\times 10^{36}$\,W & $5.81\times 10^{36}$\,W\\
\hline
\end{tabular}

\end{table}
\begin{table}

\caption{The relative contribution of the diffuse optical/NIR radiation 
$f^{opt}_{FIR}$ ($4000-22000$\,\AA),
 diffuse UV radiation $f^{UV}_{FIR}$ ($912-4000$\,\AA) and UV photons locally 
absorbed in HII regions $f^{HII}_{FIR}$ to various FIR wavelengths. The 
wavelengths considered here were those commonly used by IRAS, ISO and SCUBA,
namely 60, 100, 170, 450, and 850 ${\mu}$m. }

\vspace{0.5cm}

\begin{tabular}{|r|r|r|r|r|r|r|}
\hline\hline
\multicolumn{1}{|c}{} & \multicolumn{3}{|c}{standard model} &
\multicolumn{3}{|c|}{two-dust-disk model}\\
\multicolumn{1}{|c}{} & \multicolumn{3}{|c}{${\rm SFR}=3.5\,{\rm M}_{\odot}$/yr} & 
\multicolumn{3}{|c|}{${\rm SFR}=3.8\,{\rm M}_{\odot}$/yr}\\
\multicolumn{1}{|c}{} & \multicolumn{3}{|c}{$F=0.28$} & 
\multicolumn{3}{|c|}{$F=0.22$}\\
\hline
& & & & & &\\
${\lambda}$ & $f^{opt}_{FIR}$ & $f^{UV}_{FIR}$ & $f^{HII}_{FIR}$ & $f^{opt}_{FIR}$ & 
$f^{UV}_{FIR}$ & $f^{HII}_{FIR}$ \\
${\mu}$m  &  &      &    &  &       &    \\
& & & & & &\\
\hline
& & & & & &\\
 60 & 0.15 & 0.19 & 0.67 & 0.20 & 0.19 & 0.62 \\ 
100 & 0.24 & 0.35 & 0.41 & 0.33 & 0.37 & 0.30 \\
170 & 0.30 & 0.50 & 0.20 & 0.39 & 0.50 & 0.11 \\
450 & 0.33 & 0.60 & 0.08 & 0.39 & 0.58 & 0.03 \\
850 & 0.33 & 0.61 & 0.06 & 0.39 & 0.59 & 0.02 \\
\hline
\end{tabular}

\end{table}

We note that, regardless of the optical 
thickness of the adopted model, the diffuse optical radiation field makes 
only a relatively small contribution to the total emitted dust luminosity.
This is in qualitative agreement with various statistical inferences linking
FIR emission with young stellar populations, in particular the FIR-radio
correlation which, also after normalisation to indicators of 
galactic mass, such as K-band flux, remains rather tight (Xu et al. 1994).
Perhaps more surprising, at first glance, is the predicted predominance of 
UV-powered grain emission in the sub-mm range in both our dust 
models (Fig. 8b,d.). However, our analysis has demonstrated that the FIR 
colours have to be fundamentally interpreted also in geometrical terms, 
rather than simply as separate temperature components. Thus, the 
predicted increasing predominance of re-radiated diffuse UV photons over 
re-radiated diffuse optical photons going from the FIR into the sub-mm regime 
is due to
an increasing fraction of the emission arising from more optically
thin regions, where the ratio of optical to UV absorption is 
largely controlled by the relative optical and UV grain emissivities.
This particularly applies in the the outer disk, which, in addition to being
(relative to the inner disk) optically thin, has systematically 
lower grain heating and temperature.

\section{Summary and Conclusions}

In this paper we have modeled the edge-on spiral galaxy NGC891 from the UV to 
the FIR and
sub-mm with the purpose of understanding what is the origin of the
FIR/sub-mm emission. The strategy of the paper was to use a model for the
intrinsic distribution of older stars and associated dust as derived from 
fitting the optical and NIR images of NGC891 by Xilouris et al. (1999), 
which, supplemented by a distribution of newly-form stars, should predict the 
diffuse FIR emission. By comparing the prediction of 
the ``standard model'' with the 
observations we tried to identify whether this 
model can reproduce the observed dust emission.

To calculate the FIR emission in the galaxy we needed to have a dust model and
also the energy density of the radiation field at each position in the
galaxy. In describing the properties of the grains we used the extinction
efficiencies from Laor \& Draine (1993) for spherical \lq\lq astronomical
grains\lq\lq, for a mixture of graphites and silicates of the MRN grain size
distribution.  For small grains not in equilibrium with the radiation it was
necessary to model the stochastic emission. For this calculation we adopted the
heat capacities from Guhathakurta \& Draine (1989) for silicate grains and from
Dwek (1986) for graphite grains. Our theoretical dust model was found to be 
consistent with the extinction coefficients in the optical and NIR spectral 
range, derived independently by  Xilouris et al. (1999) for NGC891. The 
similarity between the extinction coefficients derived from our theoretical 
model and those derived as fitting parameters assure us that we can use the 
model for the distribution of old stars and associated dust of 
Xilouris et al. (1999). 

The calculation of the energy density in the galaxy was done using 
the solution of the radiation transfer equation, based on the method of 
Kylafis \& Bahcall (1987). This calculation was done separately for the
optical/NIR and UV radiation field. For the distribution of old and
intermediate age stars we used a smooth exponential disk and a de Vaucouleurs
bulge, with the parameters derived from Xilouris et al. (1999). For the
distribution of newly-formed stars we used a smooth exponential disk, with
fixed parameters, e.g., with the same scalelengths as that of the stars
radiating in the B band, and with a scaleheight of 90 pc. The
amplitude of the radiation emitted by the young stars (of the ionising and
non-ionising UV radiation) was considered as a free parameter. 
Using the population synthesis models of Mateu \& Bruzual (2000) we
parameterised  the UV radiation in terms of the recent SFR. Another free
parameter was the fraction $F$ between the UV photons which are locally absorbed
within star-forming complexes and those which participate in the diffuse
ISRF.

The predicted SED from our \lq\lq standard  model\rq\rq\, was compared
with the observed one. In the extreme case that $F=0$, the SFR can be 
taken so as to
exactly reproduce the total bolometric output in infrared. This is a solution
with SFR $=7.5\,{\rm M}_{\odot}$/yr. While requiring a high SFR, this 
solution fails
to reproduce the sub-mm part of the spectrum. A solution with $F>0$, which
includes also the contribution of star-forming complexes, underestimates
even more the emission in the sub-mm spectral range. The failure of our \lq\lq
standard model\rq\rq\, to fit the observed SED of NGC891 indicates, most 
probably, that there is missing dust, which could not have been detected by 
Xilouris et al. (1999). Two scenarios are proposed to explain this missing 
dust component, namely a scenario in which the missing dust is in the form of 
very small optically thick clumps, such that they do not affect the optical 
extinction derived by Xilouris et al. (1999), or in the 
form of a second diffuse 
thin disk, where the young stellar population is embedded. For the second 
case we give a quantitative description of the model, and we make calculations
for the FIR output. We have found that the two-dust-disk model reproduces 
very well the observed SED, as well as the radial profile at 850\,$\mu$m. This
solution requires a SFR $=3.5\,{\rm M}_{\odot}$/yr and $F=0.28$. This value of 
the SFR fits better with the preconception that NGC891 is a relatively 
quiescent normal galaxy.

Potentially, a further possibility exists to explain the shortfall in
sub-mm emission predicted by our
\lq\lq standard  model\rq\rq. A higher level of sub-mm emission could be
attained from a given grain mass if the sub-mm grain emissivity were
higher than assumed in our dust model. This approach was followed by Alton et
al. (2000a). However, for a given
grain heating the sub-mm/FIR colour of the grain emission will become colder,
in order to preserve the energy balance between the absorbed UV/optical
and the radiated emission. Thus, the peak of the SED would be shifted to longer
wavelengths, so in this scenario there would be a need to increase the SFR to
account for the observed fluxes in the 60-100\,$\mu$m range.

We emphasise that, on the basis of currently available observational 
evidence, it is difficult to distinguish between the two dust-disk-model and
the clumpy scenario. As present day sub-mm and FIR telescopes generally
have insufficient angular resolution to resolve the postulated geometrical
emission components, the best way of distinguishing between them might be to 
compare the statistics of quantities derived from the UV/optical/FIR/sub-mm 
SEDs. Such statistics could be, for example, the derived star-formation rates 
from a sample of edge-on and face-on systems with comparable morphological 
types. With the advent of the future generations of FIR telescopes with arcsec 
resolution we will be able to directly distinguish between diffuse disk and 
cloud  scenarios. Then such techniques will be needed to understand young 
galaxies at cosmological distances.

\acknowledgements

We would like to thank the anonymous referee for his useful comments and
suggestions. 
We kindly acknowledge Dr. Juan Mateu for providing us with unpublished 
results on the population synthesis models. We would also like to thank
Dr. Emmanuel Xilouris for motivating discussions.  
This work was supported in part by Projects 50OR99140 and 50QI92014 of the
Deutsches Zentrum f\"ur Luft- und Raumfahrt.

\end{document}